\documentclass[aps,nofootinbib,twocolumn,longbibliography]{revtex4-1}

\usepackage[CJKbookmarks, pdftex, bookmarksnumbered, bookmarksopen, colorlinks, citecolor=blue, linkcolor=blue]{hyperref}
\usepackage{amsmath,amssymb}
\usepackage{graphicx}

\def\id{1\hspace{-.9mm}\mathrm{l}}

\newcommand{\sectionline}{
  \nointerlineskip \vspace{\baselineskip}
  \hspace{\fill}\rule{0.5\linewidth}{.7pt}\hspace{\fill}
  \par\nointerlineskip \vspace{\baselineskip}
}

\def\be{\begin{equation}}
\def\ee{\end{equation}}

\begin{document}

\title{The boundary is mixed}

\author{Eugenio Bianchi}
\affiliation{Perimeter Institute for Theoretical Physics, 31 Caroline St.N., Waterloo ON, N2J 2Y5, Canada}

\author{Hal~M.~Haggard, Carlo Rovelli}%
\affiliation{CPT, CNRS UMR7332, Aix-Marseille Universit\'e and Universit\'e de Toulon, F-13288 Marseille, EU}

\begin{abstract} 

\noindent
We show that Oeckl's  boundary formalism incorporates quantum statistical mechanics naturally, and we formulate general-covariant quantum statistical mechanics in this language. We illustrate the formalism by showing how it accounts for the Unruh effect.  We observe that the distinction between pure and mixed states weakens in the general covariant context, and surmise that local gravitational processes are indivisibly statistical with no possible quantal versus probabilistic distinction. 

\end{abstract}

% \pacs{04.60.Pp}
% Loop Quantum Gravity

\date{\today}

\maketitle

\section{Introduction}

Quantum field theory and quantum statistical mechanics provide a framework within which most of current fundamental physics can be understood. In their usual formulation, however, they are not at ease in dealing with gravitational physics. The difficulty stems from  general covariance and the peculiar way in which general relativistic theories deal with time evolution.   A  quantum statistical  theory including gravity requires a generalized formulation of quantum and statistical mechanics.  

A key tool in this direction which has proved effective in quantum gravity, is Oeckl's idea  \cite{Oeckl:2003vu,Oeckl:2005bv} of using a boundary formalism, reviewed below. This formalism combines the advantages of an $S$-matrix transition-amplitude language with the possibility of defining the theory without referring to asymptotic regions. It is a language adapted to general covariant theories, where ``bulk" observables are notoriously tricky, because it can treat dependent and independent variables on the same footing.  This formalism allows a general covariant definition of transition amplitudes, $n$-point functions and in particular the graviton propagator \cite{Rovelli:2005yj,Bianchi:2006uf}. These are defined on compact spacetime regions---the dependence on the boundary metric data makes general covariance explicit and circumvents the difficulties  (e.g.\,\cite{ArkaniHamed:2007ky})  usually associated to the definition of these quantities  in a general covariant theory. 

In the boundary formalism, the focus is moved from ``states", which describe a system at some given time, to ``processes", which describe what happens to a local system during a finite time-span.   For a conventional non-relativistic system, the quantum space of the processes, $\cal B$ (for ``boundary"),  is simply the tensor product of the initial and final Hilbert state spaces. Tensor states in $\cal B$ represent processes with given initial and final states.  

What about the vectors in $\cal B$  that are \emph{not} of the tensor form? Remarkably, it turns out that mixed \emph{statistical} quantum states are naturally represented by these non-tensor states \cite{BianchiStoccolma}.  Here we formalize this observation, showing how statistical expectation values are expressed in this language.  This opens the way to a systematic treatment of general-covariant quantum statistical mechanics, a problem still wide open. 

The structure of this paper is as follows:  In Section \ref{II},  we start from conventional non-relativistic mechanics and move ``upward" towards more covariance: we construct the formal structures that define the boundary formalism, characterize  physical states and operators,  define the dynamics through amplitudes, and show how statistical states and equilibrium states can be treated.   In Section \ref{III}, we adapt the boundary formalism to a general covariant language by including the independent evolution parameter (the ``time" partial observable) into the configuration space. This is the step that permits the generalization to general covariant systems. Once these structures are clear, in Section \ref{IV} we take them as fundamental, and show that they retain their meaning also in the more general cases where the system is genuinely general relativistic. In Section  \ref{V} we apply the formalism to the Unruh effect and in Section \ref{VI} we draw some tentative conclusions regarding quantum gravity.  

These point towards the idea that any local gravitational process is statistical. 

\section{Non-relativistic formalism}\label{II}

\subsection{Mechanics}

Consider a Hamiltonian system with configuration space $\cal C$.  Call $x\in{\cal C}$ a generic point in ${\cal C}$.  The corresponding quantum system is defined by a Hilbert space $\cal H$ and a Hamiltonian operator $H$.  We indicate by $A,B,...\in {\cal A}$ the self-adjoint operators representing observables.  In the Schr\"odinger representation, which diagonalizes configuration variables, a state $\psi$ is represented by the functions $\psi(x)=\langle x | \psi \rangle$, where $ |x \rangle$ is a (possibly generalized) eigenvector of a family of observables that coordinatizes $\cal C$ (we use the Dirac notation also for generalized states, as Dirac did).  States evolve in time by $\psi_t=e^{-iHt}\psi_o$. For convenience we call ${\cal H}_t$  the Hilbert space isomorphic to ${\cal H}$,  thought of as the space of states at time $t$. 

Fix a time $t$ and consider the non-relativistic boundary space 
\be
      {\cal B}_t= {\cal H}_0 \otimes {\cal H}_t^*, 
\ee
where the star indicates the dual space.  This space can be interpreted as the space of all  (kinematical) \emph{processes}.  The state $\Psi= \psi \otimes\, \phi^*\in{\cal B}_t$ represents the process that takes the initial state $\psi$ into the final state $\phi$ in a time $t$.   For instance, if $\psi$ and $\phi$ are eigenstates of operators corresponding to given eigenvalues, then $\Psi$ represents a process where these eigenvalues have been measured at initial and final time.   

In the Schr\"odinger representation, vectors in $ {\cal B}_t$ have the form $\psi(x,x')=\langle x, x' | \psi \rangle$.  The state $ |x,x' \rangle\equiv  | x \rangle \otimes \langle x' |$ represents the process that takes the system from $x$ to $x'$ in a time $t$.  The interpretation of the states in  ${\cal B}_t$ which are not of the tensor form is our main concern in this paper and is discussed below. 

There are two notable structures on the space  ${\cal B}_t$.  
\begin{enumerate}
\item[(a)]  A linear function $W_t$ on ${\cal B}_t$, which  completely codes the dynamics.   This is defined by its action 
\be
         W_t(\psi \otimes\, \phi^*) :=  \langle \phi |e^{-iHt}| \psi  \rangle 
    \label{defWt}
\ee
on tensor states, and extended by linearity to the entire space. This function codes the dynamics because its value on any tensor state $\psi\otimes\phi^*$ gives the probability amplitude of the corresponding process that transforms the state $\psi$ into the state $\phi$.   Notice that the expression of $W_t$ in the Schr\"odinger basis reads
\be
         W_t(x,x') =  \langle x' |e^{-iHt}| x  \rangle,
\ee
which is precisely the Schr\"odinger-equation propagator, and can be represented formally as a Feynman path integral from $x$ to $x'$ in a time $t$, and, of course, it codes the dynamics of the theory. 

\item[(b)] There is a nonlinear map $\sigma$ that sends $\cal H$ into ${\cal B}_t$, given by 
\be
      \sigma:  \psi\mapsto  \psi  \otimes \,(e^{-iHt}\psi)^*.  
       \label{phys1}
\ee
Boundary states in the image of  $\sigma$ represent processes that have probability amplitude equal to one, as can be easily verified using \eqref{defWt} and \eqref{phys1}. The process $\sigma(\psi)$ is the one induced by the initial state $\psi$.   In general, we shall call any vector $\Psi\in{\cal B}_t$ that satisfies 
\be
      W_t(\Psi)= 1
      \label{WtPsi}
\ee
a ``physical boundary state." 

\end{enumerate}
These are the basic structures of the boundary formalism in the case of a non-relativistic system.

\subsection{Statistical mechanics}

The last equation of the previous section is linear, hence a linear combination of solutions is also a solution.  But linear combinations of tensor states are not tensor states. What do the solutions of  \eqref{WtPsi} which are \emph{not} of the tensor form represent?  

Consider a statistical state $\rho$. By this we mean here a trace class operator in $\cal H$ that can be mixed or pure. An operator in $\cal H$ is naturally identified with a \emph{vector} in ${\cal H}\otimes {\cal H}^*$, of course.  In particular, let $|n\rangle$ be an orthogonal basis that diagonalizes $\rho$, then
\be
       \rho=\sum_n\ c_n\  |n\rangle\;  \langle n |. 
       \label{rho}
\ee
The corresponding element in  ${\cal B}_0$ is 
\be
       \rho=\sum_n\ c_n\  |n\rangle \otimes  \langle n |
\ee
and we will from now on identify the two quantities. That is, below we often write states in ${\cal H}\otimes {\cal H}^*$, as operators in
$\cal H$.  The numbers $c_n$ in \eqref{rho} are the statistical weights. They satisfy 
\be
      \sum_n\ \ c_n =1
\ee
because of the trace condition on $\rho$, which expresses the fact that probabilities add up to one.  Thus the state $\rho$ can be seen as an element of ${\cal B}_0$. Consider the corresponding element of ${\cal B}_t$, defined by
\be
       \rho_t : =  \sum_n\ c_n\  |n\rangle\; \langle n |e^{iHt}. 
\ee
It is immediate to see that 
\be
      W_t(\rho_t)= 1.
      \label{Wtrhot}
\ee
Therefore we have found the physical meaning of the other (normalized) solutions of  \eqref{WtPsi}. They represent statistical states.  Notice that these are expressed as \emph{vectors} in the boundary Hilbert space ${\cal B}_t$. (See also \cite{Oeckl:2012}.)

The expectation value of the observable $A$ in the statistical state $\rho$ is 
\be
           \langle A \rangle= {\rm Tr}[A\rho],
\ee
the correlation between two observables is 
\be
           \langle AB \rangle= {\rm Tr}[AB\rho],
\ee
and the time dependent correlation is
\be
           \langle A(t)B(0) \rangle= {\rm Tr}[e^{iHt}Ae^{-iHt}B\rho],
\ee
of which the two previous expressions are special cases. These quantities can be expressed in the simple form
\be
           \langle A(t)B(0) \rangle= W_t(\ (B\otimes A)\ \rho_t)
           \label{corre}
\ee
because 
\be
           W_t((B\otimes A)\rho_t)={\rm Tr}[e^{-iHt}B\rho_tA]={\rm Tr}[e^{-iHt}B\rho e^{iHt}A],\nonumber
\ee
here the placement of $\rho_t$ within the trace reflects the fact that its left factor is in the initial space and its right factor is in the final space (and $A$ does not need a dagger because it is self-adjoint).
Therefore the boundary formalism permits a direct reformulation of quantum statistical mechanics in terms of general boundary states, boundary operators and the $W_t$ amplitude. 

Consider states of Gibbs's form $\rho =N e^{-\beta H}$.  The corresponding state in ${\cal B}_t$ is  
\be
       \rho_t =N \sum_n\  e^{-\beta E_n}  e^{iE_nt}\ |n\rangle \langle n |= N e^{iH(t+i\beta)}
\ee
where $|n\rangle$ is the energy eigenbasis and $N=N(\beta)$, determined by the normalization, is the inverse of the partition function. A straightforward calculation shows that for these states the correlations \eqref{corre} satisfy the KMS condition
\be
\langle A(t)B(0) \rangle= \langle B(-t-i\beta)A(0) \rangle
\ee
which is the mark of an equilibrium state. Thus Gibbs states are the equilibrium states. 

\subsection{$L_1$ and $L_2$ norms: physical states and pure states}

The two classes of solutions illustrated in the previous two subsections (pure states and statistical states) \textit{exhaust all solutions} of the physical boundary state condition when ${\cal B}_t$ decomposes as a tensor product of two Hilbert spaces:
\be
      {\cal B}_t= {\cal H}_0 \otimes {\cal H}_t^*. 
\ee
This can be shown as follows. Consider an orthonormal basis $| n \rangle$ in ${\cal H}_0$. Due to the unitarity of the time evolution, the vectors $( e^{-iHt} | n \rangle )^{*}$ form a basis of ${\cal H}_t^{*}$.  Therefore any state in ${\cal B}_t$ can be written in the form
\be
\Psi = \sum_{nn'} c_{nn'}  |n\rangle \otimes ( e^{-iHt} | n' \rangle )^{*}.
\ee
The physical states satisfy
\be
\langle W|\Psi\rangle= \sum_{nn'}c_{nn'}  \langle  n'|e^{iHt} e^{-iHt} |n \rangle =\sum_{n} c_{nn}=1, 
\label{L1}
\ee 
therefore they correspond precisely to the operators 
\be
\rho= \sum_{nn'} c_{nn'} |n\rangle  \langle n' |
\ee 
in ${\cal H}_0$, satisfying the condition
\be
   {\rm Tr}\rho=1
\ee
which is to say: they are the statistical states.  In particular, they are pure states if they are projection operators, $\rho^2=\rho$.

Observe that in general a statistical state in ${\cal B}_t$ is not a normalized state in this space. Rather, its $L_2$ norm satisfies 
\be
|\Psi|^2=\sum_{nn'} |c_{nn'}|^2 \le 1
\ee
where the equality holds only if the state is pure. This is easy to see in a basis that diagonalizes $\rho$, because the trace condition implies that all eigenvalues are equal or smaller than 1 and sum to 1. 

Thus there is a simple characterization of physical states and pure states: the first have the ``$L_1$" norm \eqref{L1} equal to unity.  The second have also  the ``$L_2$" norm $|\Psi|^2$ equal to unity. 

\section{Relativistic formalism}\label{III}

\subsection{Relativistic mechanics}

Let us now take a step towards the relativistic formalism where the time variable is treated on the same footing as the configuration variables.  

With this aim, consider again the same system as before and define the \emph{extended} configuration space ${\cal E}=\cal C\times \mathbb{R}$.  Call $(x,t)\in{\cal E}$ a generic point in ${\cal E}$.  Let $\Gamma_{ex}=T^{*}{\cal E}$ be the corresponding extended phase space and $C=p_t+H$ the Hamiltonian constraint, where $p_t$ is the momentum conjugate to $t$.  The corresponding quantum system is characterized by the \emph{extended} Hilbert space $\cal K$ and a Wheeler-deWitt operator $C$  \cite{Rovelli:2004fk}.  

Indicate by $A,B,...\in {\cal A}$ the self-adjoint operators representing partial observables \cite{Rovelli:2001bz} defined in $\cal K$.  In the Schr\"odinger representation that diagonalizes extended configuration variables, states are given by functions $\psi(x,t)=\langle x,t | \psi \rangle$.  The physical states are the solutions of the Wheeler-deWitt equation $C\psi=0$, which here is just the Schr\"odinger equation.   \emph{Physical} states are the (generalized) vectors $\psi(x,t)$ in $\cal K$ that are solutions of the Schr\"odinger equation. 

The space $\cal H$ formed by the physical states that are solutions of the Schr\"odinger equation is clearly in one-to-one correspondence with the space ${\cal H}_0$ of the states at time $t\!=\!0$. Therefore there is a linear map that sends ${\cal H}_0$ into (a suitable completion of) $\cal K$, simply defined by sending the state $\psi(x)$ into the solution $\psi(x,t)$ of the Schr\"odinger equation such that  $\psi(x,0)=\psi(x)$. Vice versa, there is a (generalized) projection $P$ from (a dense subspace of) $\cal K$ to $\cal H$, that sends a state $\psi(x,t)$ to a solution of the Schr\"odinger  equation. This can be formally obtained from the spectral decomposition of $C$, or, more simply, by 
\be
          (P\psi)(x,t)=\int dx'\, dt'\  W_{(t-t')}(x,x')\ \psi(x',t').
\ee

Now, \emph{without} fixing a time, the \emph{relativistic} boundary state space is defined by 
\be
      {\cal B}={\cal K}\otimes {\cal K}^*. 
\ee
Notice the absence of the $t$-label subscript. In the Schr\"odinger representation, vectors in ${\cal B}$ have the form $\psi(x,t,x',t')=\langle x, t, x', t' | \psi \rangle$.  This space can again be interpreted as the space of all (kinematical) \emph{processes}, where now the boundary measurement of the clock time $t$ is treated on the same footing as the other partial observables. Thus for instance  $| x, t, x', t' \rangle\equiv| x, t \rangle\otimes \langle x',t'|$ represents the process that takes the system from the configuration $x$ at time $t$ to the configuration $x'$ at time $t'$.

The two structures considered above simplify on the space  ${\cal B}$.  
\begin{enumerate}
\item[(a)] The dynamics is completely coded by a linear function $W$ (no $t$ label!) on $\cal B$. This is defined extending by linearity  
\be
         W(\phi^*\otimes\psi) :=  \langle \phi |P| \psi  \rangle.
\ee
Its expression in the Schr\"odinger basis reads
\be
\begin{aligned}
         W(x,t,x',t') &=  \langle x, t |P| x',t'  \rangle \\
         &= \langle x|e^{iH(t-t')}| x'\rangle, 
         \label{defW}
\end{aligned}
\ee
which is once again nothing but the  Schr\"odinger-equation propagator, now seen as a function of initial and final \emph{extended} configuration variables.  The  variable $t$ is not treated as an independent evolution parameter, but rather is treated on equal footing with the other partial observables.  The operator $P$ can still be represented as a suitable Feynman path integral in the extended configuration space, from the point $(x,t)$ to the point $(x',t')$. 

\item[(b)] Second, there is again a nonlinear map $\sigma$ that sends $\cal K$ into ${\cal B}$, now simply given by 
\be
        \sigma: \psi\mapsto \psi\ \otimes \ \psi^*. 
       \label{phys}
\ee
States in the image of this map are ``physical", namely represent processes that have probability amplitude equal to one, only if $\psi$ satisfies the  Schr\"odinger equation. In this case, a straightforward calculation verifies that 
\be
      W(\Psi)= 1. 
      \label{WPsi}
\ee
As before, we call ``physical" any state in $\cal B$ solving this equation. 
\end{enumerate}

\subsection{Relativistic statistical mechanics}

As before, linear combinations of physical states represent statistical states.  A general relativistic statistical state is a statistical superposition of solutions of the equations of motion \cite{Rovelli:1993ys}.\footnote{A concrete example is illustrated in \cite{Rovelli:1993zz}.}  Consider again the state \eqref{rho} in this language: if  $\psi_n$ is the full time-dependent solution of the Schr\"odinger equation corresponding to the initial state $|n\rangle$, we can now represente the state  \eqref{rho}  in $\cal B$ simply by 
\be
       \rho=\sum_n\ c_n\  \psi_n\,  \psi_n^*. 
\ee
Explicitly, in the  Schr\"odinger basis
\be
       \rho(x,t,x',t')=\sum_n\ c_n\  \psi_n(x,t)\ \overline{\psi_n(x',t')}. 
\ee

The equilibrium statistical state at inverse temperature $\beta$ is given by 
\begin{eqnarray}
        \rho(x,t,x',t')&=& N \sum_n\   e^{iE_n (t-t'+i\beta)}\ \psi_n(x)\ \overline{\psi_n(x')}. \nonumber \\
        &=& N\ e^{i H (t-t'+i\beta)}.
\end{eqnarray}
where $\psi_n(x)$ are the energy eigenfunctions.

The correlation functions between partial observables are now given simply by 
\be
           \langle AB \rangle= W((A\otimes B)\ \rho).
           \label{correr}
\ee
Notice the complete absence of the time label $t$ in the formalism. Any temporal dependence is folded into the boundary data. (However, see the next section for a generalization of the KMS property and equilibrium.) 

This completes the construction of the boundary formalism for a relativistic system. We now have at our disposal the full language and we can ``throw away the ladder," keep only the structure constructed, and extend it to far more arbitrary systems, including relativistic gravity. 

\section{General boundary}\label{IV}

We now generalize the boundary formalism to genuinely (general) relativistic systems that do not have a non-relativistic formulation. 

A quantum system is defined by the triple $({\cal B},{\cal A},{W})$. The Hilbert space $\cal B$ is interpreted as the boundary state space, not necessarily of the tensor form.  $\cal A$ is an algebra of self-adjoint operators on $\cal B$. The elements $A,B,...\in\cal A$ represent partial observables, namely quantities to which we can imagine associating measurement apparatuses, but whose outcome is not necessarily predictable (think for instance of a clock). The linear map $W$ on $\cal B$ defines the dynamics.   

Vectors $\Psi\in{\cal B}$ represent processes. If $\Psi$ is an eigenstate of the operator $A\in{\cal A}$ with eigenvalue $a$, it represents a process where the corresponding boundary observable has value $a$.  The quantity 
\be
    W(\Psi)  \:= \ \langle W| \Psi \rangle 
\ee
is the amplitude of the process.  Its modulus square (suitably normalized) determines the relative probability of  distinct processes \cite{Rovelli:2004fk}.
A \emph{physical} process is a vector in $\cal B$ that has amplitude equal to one, namely satisfies 
\be
    \langle W| \Psi \rangle=1. 
\ee
The expectation value of an operator $A\in{\cal A}$ on a physical process $\Psi$ is 
\be
  \langle A \rangle =    \langle W |  A  | \Psi \rangle. 
\ee

If a tensor structure in $\cal B$ is not given, then there is no \emph{a priori\ }Êdistinction between pure and mixed states.  The distinction between quantum incertitude and statistical incertitude acquires meaning only if we can distinguish past and future parts of the boundary \cite{Smolin:1982wt,Smolin:1986hv}.  

So far, there is no notion of time flow in the theory.  The theory predicts correlations between boundary observables. However, as pointed out in \cite{Connes:1994hv}, a generic state $\Psi$ on the algebra of local observables of a region defines a flow $\alpha_\tau$ on the observable algebra by the Tomita theorem \cite{Connes:1994hv}, and the state $\Psi$  satisfies the KMS condition for this flow
\be
\langle A(\tau)B(0) \rangle= \langle B(-\tau-i\beta)A(0) \rangle,
\ee
where $A(\tau)=\alpha_\tau(A)$.  It will be interesting to compare the flow generated in this manner with the flow generated by a statistical state within the boundary Hilbert space. 

If a flow is given a priori, the KMS states for this flow are equilibrium states for this flow.  

In a general relativistic theory including gravity, no flow is given a priori, but we can still distinguish physical equilibrium states as follows: an equilibrium state is a state that defines a mean geometry and whose Tomita flow is given by a timelike Killing vector of this geometry: see \cite{Rovelli:2012nv}. \\

\section{Unruh effect}\label{V}

As an example application of the formalism, we describe the Unruh effect \cite{Unruh:1976db} in this language. Other treatments with a focus on the general boundary formalism are \cite{Colosi:2013,Hellman:2012}. Consider a partition of Minkowski space into two regions $M$ and $\tilde M$ separated by the two surfaces
\be
           \Sigma_0: \{t=0, \ x\ge0\},
\hspace{1em}
           \Sigma_\eta: \{t=\eta x,\   x\ge0\}. 
\ee 
The region $M$ is a wedge of angular opening $\eta$ and $\tilde M$ is its complement  (Figure 1). 
\begin{figure}[h]
\includegraphics[width=5.5cm]{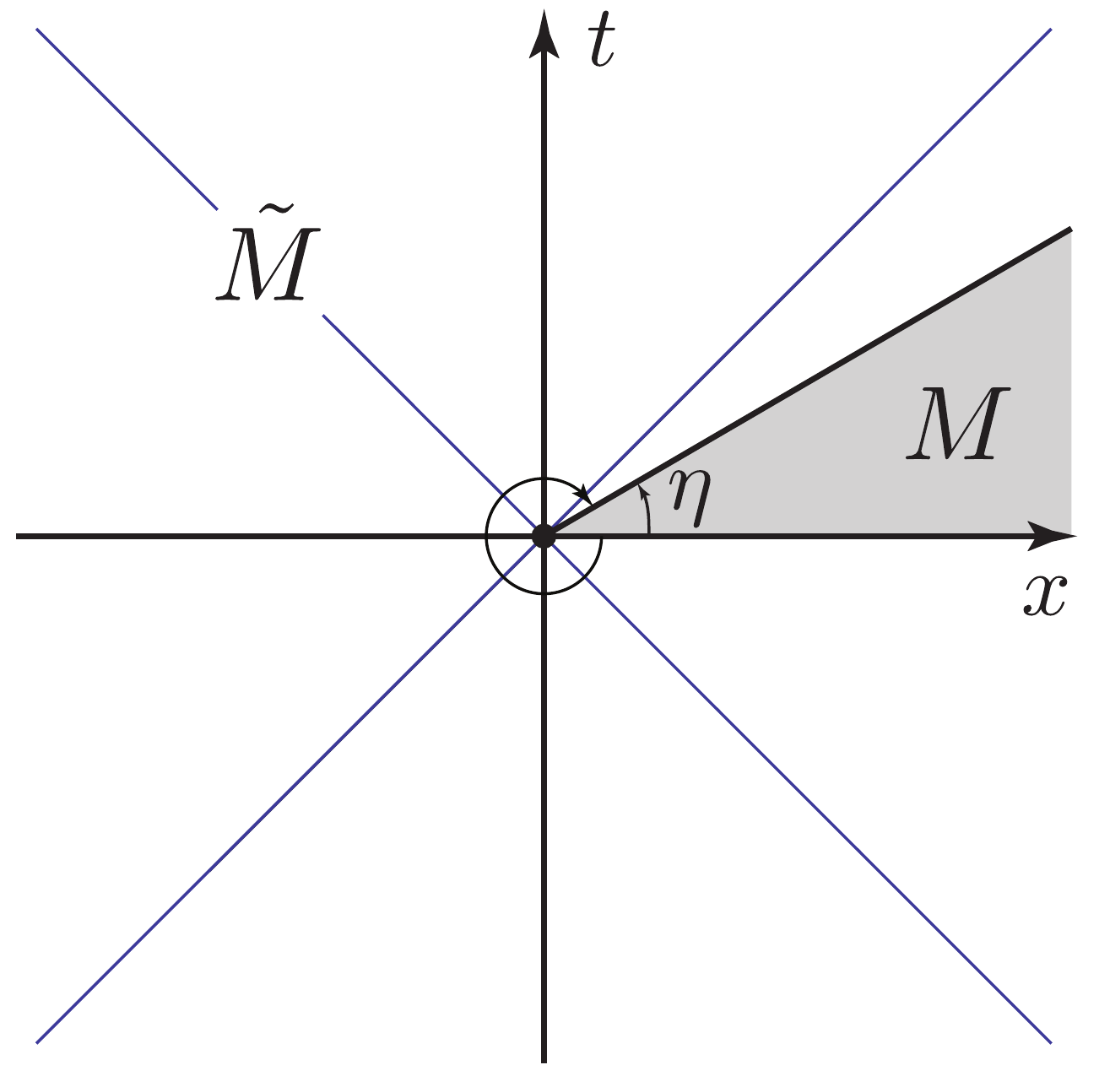}
\caption{The wedge $M$ in Minkowski space.}
\end{figure} 
Consider a Lorentz invariant quantum field theory on Minkowski space, say satisfying the Wigtmam axioms   \cite{Streater:2000fk}: in particular, energy is positive-definite and there is a single Poincar\'e-invariant state, the  vacuum $|0\rangle$. How is the vacuum described in the boundary language?  

In general,  a boundary state $\phi_b$ on $\partial M=\Sigma=\Sigma_0\cup\Sigma_\eta$ is a vector in the Hilbert space ${\cal B}={\cal H}_0\otimes {\cal H}_\eta^*$, where ${\cal H}_0$ and ${\cal H}_\eta$ are Hilbert spaces associated to the states on $\Sigma_0$ and $\Sigma_\eta$ respectively.  The conventional Hilbert space ${\cal H}$ associated to the $t\!=\!0$ surface is the tensor product of two Hilbert spaces ${\cal H}={\cal H}_L\otimes{\cal H}_R$ that describe the degrees of freedom to the left or right of the origin. We can identify ${\cal H}_R$ and ${\cal H}_0$ since they carry the same observables: the field operators on $\Sigma_0$.  Because the theory is Lorentz-invariant, ${\cal H}$ carries a representation of the Lorentz group. The self-adjoint boost generator $K$ in the $t,x$ plane does not mix the two factors ${\cal H}_L$ and ${\cal H}_R$. If we call $k$ its eigenvalues and $|k, \alpha \rangle_{L,R}$, its eigenstates in the two factors with $\alpha$ labeling the distinct degenerate levels of $k$, then it is a well known result \cite{Bisognano:1976za} that 
\be
\langle 0|k, \alpha \rangle_L=e^{-\pi k}  \langle k, \alpha|_R 
\ee
which we can write in the form 
\be
|0\rangle = \int dk\, d\alpha\, e^{-\pi k}\,|k,\alpha\rangle_L \otimes|k,\alpha\rangle_R.  
\ee
Tracing over ${\cal H}_L$ gives the density matrix in ${\cal H}_R$
\be
\rho_0= {\rm Tr}_L\big[|0\rangle\langle 0|\big] = e^{-2\pi K}
\ee
which determines the result of any vacuum measurement, and therefore \emph{any} measurment \cite{Wightman:1959fk}, performed on $\Sigma_0$. The evolution operator $W_\eta$ in the angle $\eta$, associated to the wedge, sends $\Sigma_0$ to $\Sigma_\eta$ and is 
\be
W_{\eta}=e^{-i \eta K}.
\ee
These two quantities give immediately the boundary expression of the vacuum on $\Sigma$:
\be
\rho_\eta=  \rho_0 e^{i\eta K} = e^{i(\eta + 2\pi i) K}
\ee
This is the vacuum in the boundary formalism. It is a KMS state at temperature $1/2\pi$ with respect to the flow generated by $K$ in $\eta$.  For an observer moving with constant proper acceleration $a$ along the hyperboloid of points with constant distance from the origin, this flow is proportional to proper time $s$ 
\be
         s=\eta/a.
\ee
And therefore  the vacuum is a KMS state, namely a thermal state, at the Unruh temperature (restoring $\hbar$)
\be
         T=\frac{\hbar a}{2\pi}.
\ee
This is the manner in which the Unruh effect is naturally described in the boundary language.  Notice that no reference to accelerated observers or special basis in Hilbert space is needed to identify the thermal character of the vacuum on the $\eta$-wedge. 

An interesting remark is that the expectation values of operators on $\Sigma$ can be equally computed using the region $\tilde M$ which is \emph{complementary} to the wedge $M$. Let us first do this for $\eta=0$. In this case, the insertion of the empty region $M$ cannot alter the value of the observables, and therefore it is reasonable to take the boundary state we associate to it to be the unit operator.
\be
         \tilde \rho=\id
\ee
And therefore 
\be
         \tilde \rho_\eta= e^{-i \eta K}.
\ee
For consistency, we have then that the evolution operator associated to $\tilde M$ must be
\be
\tilde W_\eta= e^{i(\eta + 2\pi i) K}.
\label{2pi}
\ee
Therefore the evolution operator and the boundary state  simply swap their roles when going from a region to its complement.\footnote{This can be intuitively understood in terms of path integrals: the evolution operator is the path integral on the interior of a spacetime region, at fixed boundary values; the boundary state can be viewed as the path integral on the exterior of the region.  In the case under consideration, the vacuum is singled out by the boundary values of the field at infinity. For a detailed discussion, see \cite{Bianchi:2013fk}.}
Notice that there exists a geometrical transformation that rotates $\Sigma_0$ into $\Sigma_\eta$, obtained by rotating it clockwise, rather than anti clockwise. This rotation is not implemented by a proper Lorentz transformation, because the Lorentz group rotates $\Sigma_0$ at most only up to the light cone $t\!=\!-x$.  But it can nevertheless be realized by extending a Lorentz transformation 
\begin{eqnarray}
x'&=&\cosh(\eta)x+\sinh(\eta)t    \nonumber \\
t'&=&\sinh(\eta)x+\cosh(\eta)t
\end{eqnarray}
to a complex parameter $i\eta$
\begin{eqnarray}
x'&=&\cosh(i\eta)x+\sinh(i\eta)t  =\cos(\eta)x+i \sin(\eta)t   \nonumber \\
t'&=&\sinh(i\eta)x+\cosh(i\eta)t = i \sin(\eta)x+\cos(\eta)t.
\end{eqnarray}
For a small $\eta=\epsilon$, this transformation rotates the positive $x$ axis infinitesimally into the complex $t$ plane. The Lorentz group acts on the expectation values of the theory, and in particular on the expectation values of products of its local observables.  Since the $n$-point functions of a quantum field theory where the energy is positive can be continued analytically for complex times (Theorem 3.5, pg. 114 in \cite{Streater:2000fk}), this action is well defined on expectation values. In particular, we can rotate $(t,x)$ infinitesimally into the complex $t$ plane, and then rotate around the real $t,x$ plane, passing below the light cone $x=\pm t$ in complex space.  In other words, by adding a small complex rotation into imaginary time, we can rotate a space-like half-line into a timelike one  \cite{Gibbons:1977ys,Bianchi:2012ui}. A full rotation is implemented by $U(2\pi i)$, giving \eqref{2pi}.   

Finally, observe that the vacuum is the \emph{unique} Poincar\'e invariant state in the theory.  This implies that if a state is Poincar\'e invariant then it is thermal at the Unruh temperature on the boundary of the wedge.     This is clearly a reflection of  correlations with physics beyond the edge of the wedge. 

Since vacuum expectation values determine all local measurable quantum-field-theory observables, this implies that the boundary state is unavoidably mixed. In essence the available field operators are insufficient to purify the state. This can be seen physically as follows: in principle, we can project the state onto a \emph{pure} state on $\Sigma_0$, breaking Poincar\'e invariance by singling out the origin, but to do so we need a complete measurement   of field values for $x>0$ and therefore an infinite number of measurements, which would move the state out of its folium \cite{Haag:1996}.  We continue these considerations in the next section.

\section{Relation with Gravity and thermality of gravitational states}\label{VI}

So far, gravity has played no direct role in our considerations.  The construction above, however, is motivated by general relativity, because the boundary formalism is not needed as long as we deal with a quantum field theory on a fixed geometry, but becomes crucial in quantum gravity, where it allows us to circumvent the difficulties raised by diffeomorphism invariance in the quantum context.   

In quantum gravity we can study probability amplitudes for local processes by associating boundary states to a finite portion of spacetime,  \emph{and including the quantum dynamics of spacetime itself in the process}.  Therefore the boundary state includes the information about the geometry of the region itself.   

The general structure of statistical mechanics of relativistic quantum geometry has been explored in \cite{Rovelli:2012nv}, where equilibrium states are characterized as those whose Tomita flow is a Killing vector of the mean geometry. Up until now it hasn't been possible to identify the statistical states in the general boundary formalism and so this strategy hasn't been available in this more covariant context. With a boundary notion of statistical states this becomes possible. It becomes possible, in particular, to check if given boundary data allow for a mean geometry that interpolates them.  

In quantum gravity we are interested in spacelike boundary states where initial and final data can be given, therefore a typical spacetime region will have the lens shape depicted in Figure 2. Past and future components of the boundary will meet on wedge-like two-dimensional ``corner" regions.  Now, say we assume that a quantum version of the equivalence principle holds, for which the local physics at the corner is locally Lorentz invariant.   Then the result of the previous section indicates that the boundary state of the lens region will be mixed.  Any such boundary state in quantum gravity is a mixed state.  (Other arguments for the thermality of local spacetime processes are in \cite{Martinetti:2002sz}.) The dynamics at the corner is governed by the corner terms of the action \cite{Carlip:1993sa,Bianchi:2012vp}, which can indeed be seen as responsible for the thermalization \cite{Massar:1999wg,Jacobson:2003wv}.  

\begin{figure}[t]
\includegraphics[height=1.5cm]{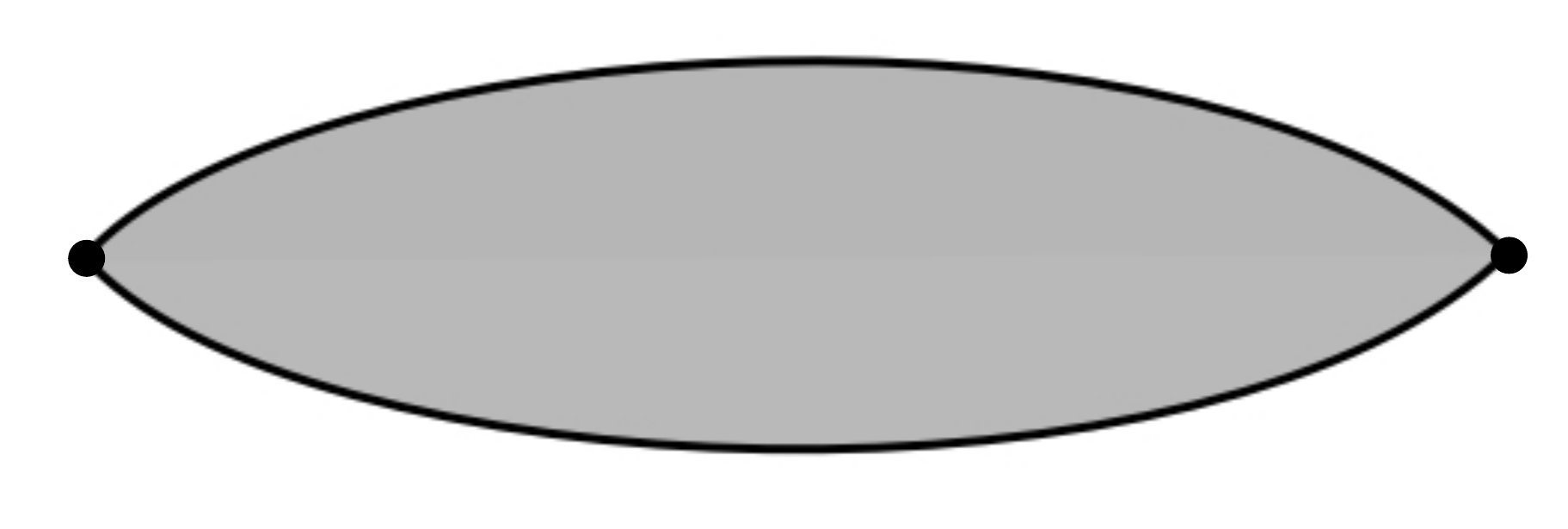}
\caption{Lens shaped spacetime region with spacelike boundaries and corners (filled circles).}
\end{figure} 

Up to this point we have emphasized the mixed state character of the boundary states in order to make a clear connection with the standard quantum formalism. However, note that from the perspective of the fully covariant general boundary formalism (see section \ref{IV}) there is always a single boundary Hilbert space ${\cal B}$ that can be made bipartite in many different manners. From this point of view it is more natural to call these boundary states non-separable. Then, local gravitational states are entangled states. This was first appreciated in the context of the examples treated in \cite{Bianchi:2013fk}, which was an inspiration for the present work. 

Recently Bianchi and Myers have conjectured that in a theory of quantum gravity, for any sufficiently  large region corresponding to a smooth background spacetime, the entanglement entropy between the degrees of freedom describing the given region with those describing its complement are given by the Bekenstein-Hawking entropy \cite{Bianchi:2012ev}. The Bianchi-Myers conjecture and the considerations above result in a compelling picture supporting a quantum version of the equivalence principle. %\footnote{This further sharpens the questions surrounding the firewall argument \cite{Almheiri:2012rt}.}  

Both the mixing of the state near a corner and the Bianchi-Myers conjecture  can be seen as manifestations of the fact that by restricting the region of interest to a finite spatial region we are tracing over the correlations between this region and the exterior, and therefore we are necessarily dealing with a state which is not pure.  If, as we expect, the boundary formalism is crucial for extracting physical amplitudes from quantum gravity, all this appears to imply that the notion of pure state is irrelevant in local quantum gravitational physics and therefore statistical fluctuations cannot be disentangled from quantum fluctuations in quantum gravity  \cite{Smolin:1982wt,Smolin:1986hv}.

\sectionline

EB acknowledges support from a Banting Postdoctoral Fellowship from NSERC. Research at Perimeter Institute is supported by the Government of Canada through Industry Canada and by the Province of Ontario through the Ministry of Research \& Innovation. HMH acknowledges support from the National Science Foundation (NSF) International Research Fellowship Program (IRFP) under Grant No. OISE-1159218.


\begin{thebibliography}{10}

\bibitem{Oeckl:2003vu}
R.~Oeckl, ``{A `general boundary' formulation for quantum mechanics and quantum
  gravity},'' \href{http://dx.doi.org/10.1016/j.physletb.2003.08.043}{{\em
  Phys. Lett.} {\bf B575} (2003)  318--324},
\href{http://arxiv.org/abs/hep-th/0306025}{{\tt arXiv:hep-th/0306025}}.
%%CITATION = HEP-TH/0306025;%%.

\bibitem{Oeckl:2005bv}
R.~Oeckl, ``{General boundary quantum field theory: Foundations and probability
  interpretation},'' {\em Adv. Theor. Math. Phys.} {\bf 12} (2008)  319--352,
\href{http://arxiv.org/abs/hep-th/0509122}{{\tt arXiv:hep-th/0509122}}.
%%CITATION = HEP-TH/0509122;%%.

\bibitem{Rovelli:2005yj}
C.~Rovelli, ``{Graviton propagator from background-independent quantum
  gravity},'' \href{http://dx.doi.org/10.1103/PhysRevLett.97.151301}{{\em Phys.
  Rev. Lett.} {\bf 97} (2006)  151301},
\href{http://arxiv.org/abs/gr-qc/0508124}{{\tt arXiv:gr-qc/0508124}}.
%%CITATION = GR-QC/0508124;%%.

\bibitem{Bianchi:2006uf}
E.~Bianchi, L.~Modesto, C.~Rovelli, and S.~Speziale, ``{Graviton propagator in
  loop quantum gravity},''
  \href{http://dx.doi.org/10.1088/0264-9381/23/23/024}{{\em Class. Quant.
  Grav.} {\bf 23} (2006)  6989--7028},
\href{http://arxiv.org/abs/gr-qc/0604044}{{\tt arXiv:gr-qc/0604044}}.
%%CITATION = GR-QC/0604044;%%.

\bibitem{ArkaniHamed:2007ky}
N.~Arkani-Hamed, S.~Dubovsky, A.~Nicolis, E.~Trincherini, and G.~Villadoro,
  ``{A Measure of de Sitter entropy and eternal inflation},''
  \href{http://dx.doi.org/10.1088/1126-6708/2007/05/055}{{\em JHEP} {\bf 0705}
  (2007)  055},
\href{http://arxiv.org/abs/0704.1814}{{\tt arXiv:0704.1814}}.
%%CITATION = ARXIV:0704.1814;%%.

\bibitem{BianchiStoccolma}
E.~Bianchi, ``Talk at the 2012 Marcel Grossmann meeting,'' July, 2012.

\bibitem{Oeckl:2012}
R.~Oeckl, ``A positive formalism for quantum theory in the general boundary formulation,"
\href{http://arxiv.org/abs/1212.5571}{{\tt arXiv:quant-ph/1212.5571}}.

\bibitem{Rovelli:2004fk}
C.~Rovelli, {\em Quantum Gravity}.
\newblock Cambridge University Press, Cambridge, U.K., 2004.

\bibitem{Rovelli:2001bz}
C.~Rovelli, ``{Partial observables},''
  \href{http://dx.doi.org/10.1103/PhysRevD.65.124013}{{\em Phys. Rev.} {\bf
  D65} (2002)  124013},
\href{http://arxiv.org/abs/gr-qc/0110035}{{\tt arXiv:gr-qc/0110035}}.
%%CITATION = GR-QC/0110035;%%.

\bibitem{Rovelli:1993ys}
C.~Rovelli, ``{Statistical mechanics of gravity and the thermodynamical origin
  of time},''
{\em Class. Quant. Grav.} {\bf 10} (1993)  1549--1566.
%%CITATION = CQGRD,10,1549;%%.

\bibitem{Rovelli:1993zz}
C.~Rovelli, ``{The Statistical state of the universe},''
\href{http://dx.doi.org/10.1088/0264-9381/10/8/016}{{\em Class. Quant. Grav.}
  {\bf 10} (1993)  1567}.
%%CITATION = CQGRD,10,1567;%%.

\bibitem{Smolin:1982wt}
L.~Smolin, ``On the nature of quantum fluctuations and their relation to
  gravitation and the principle of inertia,''
{\em Classical and Quantum Gravity} {\bf 3} (1986)  347.
%%CITATION = PRINT-83-0106 (IAS,PRINCETON) ETC.;%%.

\bibitem{Smolin:1986hv}
L.~Smolin, ``Quantum gravity and the statistical interpretation of quantum
  mechanics,''
\href{http://dx.doi.org/10.1007/BF00668705}{{\em Int. J. Theor. Phys.} {\bf 25}
  (1986)  215--238}.
%%CITATION = IJTPB,25,215;%%.

\bibitem{Connes:1994hv}
A.~Connes and C.~Rovelli, ``{Von Neumann algebra automorphisms and time
  thermodynamics relation in general covariant quantum theories},'' {\em Class.
  Quant. Grav.} {\bf 11} (1994)  2899--2918,
\href{http://arxiv.org/abs/gr-qc/9406019}{{\tt arXiv:gr-qc/9406019}}.
%%CITATION = GR-QC/9406019;%%.

\bibitem{Rovelli:2012nv}
C.~Rovelli, ``{General relativistic statistical mechanics},''
\href{http://arxiv.org/abs/1209.0065}{{\tt arXiv:1209.0065}}.
%%CITATION = ARXIV:1209.0065;%%.

\bibitem{Unruh:1976db}
W.~Unruh, ``{Notes on black hole evaporation},''
\href{http://dx.doi.org/10.1103/PhysRevD.14.870}{{\em Phys.Rev.} {\bf D14}
  (1976)  870}.
%%CITATION = PHRVA,D14,870;%%.

\bibitem{Colosi:2013}
D.~Colosi and D.~R{\"a}tzel, ``{The Unruh Effect in General Boundary Quantum Field Theory},"
\href{http://dx.doi.org/10.3842/SIGMA.2013.019}{{\em SIGMA} {\bf 9}
  (2013)  19}.

\bibitem{Hellman:2012}
R.~Banisch, F.~Hellman, and D.~R{\"a}tzel, ``{Vacuum states on timelike hypersurfaces in quantum field theory},"
\href{http://arxiv.org/abs/1205.1549}{{\tt arXiv:1205.1549}}.

\bibitem{Streater:2000fk}
R.~Streater and A.~Wightman, {\em PCT, Spin and Statistics, and All That}.
\newblock Princeton University Press, 2000.

\bibitem{Bisognano:1976za}
J.~Bisognano and E.~Wichmann, ``{On the Duality Condition for Quantum
  Fields},''
\href{http://dx.doi.org/10.1063/1.522898}{{\em J.Math.Phys.} {\bf 17} (1976)
  303--321}.
%%CITATION = JMAPA,17,303;%%.

\bibitem{Wightman:1959fk}
A.~Wightman, ``Quantum field theory in terms of vacuum expectation values,''
  {\em Physical Review} {\bf 101} (1959) no.~860.

\bibitem{Bianchi:2013fk}
E.~Bianchi and H.~M. Haggard, {\em to appear} (2013).

\bibitem{Gibbons:1977ys}
G.~Gibbons and S.~Hawking, ``Action integrals and partition functions in
  quantum gravity,'' \href{http://dx.doi.org/10.1103/PhysRevD.15.2752}{{\em
  Physical Review D} {\bf 15} (1977) no.~10, 2752--2756}.

\bibitem{Bianchi:2012ui}
E.~Bianchi, ``{Entropy of Non-Extremal Black Holes from Loop Gravity},''
\href{http://arxiv.org/abs/1204.5122}{{\tt arXiv:1204.5122}}.
%%CITATION = ARXIV:1204.5122;%%.

\bibitem{Haag:1996}
R.~Haag, {\em Local Quantum Physics}.
\newblock Springer-Verlag, Berlin Heidelberg New York, 1996.

\bibitem{Martinetti:2002sz}
P.~Martinetti and C.~Rovelli, ``{Diamonds's temperature: Unruh effect for
  bounded trajectories and thermal time hypothesis},''
  \href{http://dx.doi.org/10.1088/0264-9381/20/22/015}{{\em Class. Quant.
  Grav.} {\bf 20} (2003)  4919--4932},
\href{http://arxiv.org/abs/gr-qc/0212074}{{\tt arXiv:gr-qc/0212074}}.
%%CITATION = GR-QC/0212074;%%.

\bibitem{Carlip:1993sa}
S.~Carlip and C.~Teitelboim, ``{The Off-shell black hole},''
  \href{http://dx.doi.org/10.1088/0264-9381/12/7/011}{{\em Class.Quant.Grav.}
  {\bf 12} (1995)  1699--1704},
\href{http://arxiv.org/abs/gr-qc/9312002}{{\tt arXiv:gr-qc/9312002}}.
%%CITATION = GR-QC/9312002;%%.

\bibitem{Bianchi:2012vp}
E.~Bianchi and W.~Wieland, ``{Horizon energy as the boost boundary term in
  general relativity and loop gravity},''
\href{http://arxiv.org/abs/1205.5325}{{\tt arXiv:1205.5325}}.
%%CITATION = ARXIV:1205.5325;%%.

\bibitem{Massar:1999wg}
S.~Massar and R.~Parentani, ``{How the change in horizon area drives black hole
  evaporation},'' \href{http://dx.doi.org/10.1016/S0550-3213(00)00067-5}{{\em
  Nucl.Phys.} {\bf B575} (2000)  333--356},
\href{http://arxiv.org/abs/gr-qc/9903027}{{\tt arXiv:gr-qc/9903027}}.
%%CITATION = GR-QC/9903027;%%.

\bibitem{Jacobson:2003wv}
T.~Jacobson and R.~Parentani, ``{Horizon entropy},''
  \href{http://dx.doi.org/10.1023/A:1023785123428}{{\em Found.Phys.} {\bf 33}
  (2003)  323--348},
\href{http://arxiv.org/abs/gr-qc/0302099}{{\tt arXiv:gr-qc/0302099}}.
%%CITATION = GR-QC/0302099;%%.

\bibitem{Bianchi:2012ev}
E.~Bianchi and R.~C. Myers, ``{On the Architecture of Spacetime Geometry},''
\href{http://arxiv.org/abs/1212.5183}{{\tt arXiv:1212.5183}}.
%%CITATION = ARXIV:1212.5183;%%.

%\bibitem{Almheiri:2012rt}
%A.~Almheiri, D.~Marolf, J.~Polchinski, and J.~Sully, ``{Black Holes:
%  Complementarity or Firewalls?},''
%  \href{http://dx.doi.org/10.1007/JHEP02(2013)062}{{\em JHEP} {\bf 1302} (2013)
%   062},
%\href{http://arxiv.org/abs/1207.3123}{{\tt arXiv:1207.3123}}.
%%CITATION = ARXIV:1207.3123;%%.

\end{thebibliography}
\end{document}